\documentclass[nofootinbib,aps,prl,reprint,groupedaddress,superscriptaddress,showpacs]{revtex4-1}
\usepackage[english]{babel}
\usepackage[utf8x]{inputenc}
\usepackage{amssymb}
\usepackage{amsthm}
\usepackage{amsmath}
\usepackage{hyperref}
\usepackage{subfigure}
\usepackage[percent]{overpic} 
\usepackage{picture}
\usepackage{color}
\usepackage{nth}
\usepackage{gensymb}
\usepackage{cleveref}
\usepackage{textcomp}
\usepackage{makecell}
\usepackage{lineno} 
\usepackage{textcomp}
\graphicspath{{img/}}

\begin{document}

\title{Energy spread minimization in a beam-driven plasma wakefield accelerator}

\author{R. Pompili}
\email[]{riccardo.pompili@lnf.infn.it}
\author{M.P. Anania}
\author{M. Behtouei}
\author{M. Bellaveglia}
\author{A. Biagioni}
\author{F.G. Bisesto}
\affiliation{Laboratori Nazionali di Frascati, Via Enrico Fermi 54, 00044 Frascati, Italy}
\author{M. Cesarini}
\affiliation{Laboratori Nazionali di Frascati, Via Enrico Fermi 54, 00044 Frascati, Italy}
\affiliation{Sapienza University, Piazzale Aldo Moro 5, 00185 Rome, Italy}
\author{E. Chiadroni}
\affiliation{Laboratori Nazionali di Frascati, Via Enrico Fermi 54, 00044 Frascati, Italy}
\author{A. Cianchi}
\affiliation{University or Rome Tor Vergata and INFN, Via Ricerca Scientifica 1, 00133 Rome, Italy}
\author{G. Costa}
\author{M. Croia}
\author{A. Del Dotto}
\author{D. Di Giovenale}
\author{M. Diomede}
\author{F. Dipace}
\author{M. Ferrario}
\author{A. Giribono}
\author{V. Lollo}
\author{L. Magnisi}
\author{M. Marongiu}
\affiliation{Laboratori Nazionali di Frascati, Via Enrico Fermi 54, 00044 Frascati, Italy}
\author{A. Mostacci}
\affiliation{Sapienza University, Piazzale Aldo Moro 5, 00185 Rome, Italy}
\author{G. Di Pirro}
\author{S. Romeo}
\affiliation{Laboratori Nazionali di Frascati, Via Enrico Fermi 54, 00044 Frascati, Italy}
\author{A.R. Rossi}
\affiliation{INFN Milano, via Celoria 16, 20133 Milan, Italy}
\author{J. Scifo}
\author{V. Shpakov}
\author{C. Vaccarezza}
\author{F. Villa}
\affiliation{Laboratori Nazionali di Frascati, Via Enrico Fermi 54, 00044 Frascati, Italy}
\author{A. Zigler}
\affiliation{Laboratori Nazionali di Frascati, Via Enrico Fermi 54, 00044 Frascati, Italy}
\affiliation{Racah Institute of Physics, Hebrew University, 91904 Jerusalem, Israel}


\date{\today}

\begin{abstract}
Next-generation plasma-based accelerators can push electron bunches to gigaelectronvolt energies within centimetre distances~\cite{faure2006controlled,gonsalves2019petawatt}.
The plasma, excited by a \textit{driver} pulse, generates large electric fields that can efficiently accelerate a trailing \textit{witness} bunch~\cite{litos2014high,steinke2016multistage,adli2018acceleration} making possible the realization of laboratory-scale applications ranging from high-energy colliders~\cite{lee2002energy} to ultra-bright light sources~\cite{nakajima2008towards}.
So far several experiments have demonstrated a significant acceleration~\cite{mangles2004monoenergetic,2007Natur.445..741B,deng2019generation} but the resulting beam quality, especially the energy spread, is still far from state of the art conventional accelerators.
Here we show the results of a beam-driven plasma acceleration experiment where we used an
electron bunch as a driver followed by an ultra-short witness.
The experiment demonstrates, for the first time, an innovative method to achieve an ultra-low energy spread of the accelerated witness of about $0.1\%$. This is an order of magnitude smaller than what has been obtained so far.
The result can lead to a major breakthrough toward the optimization of the plasma acceleration process and its implementation in forthcoming compact machines for user-oriented applications.
\end{abstract}


\maketitle

The idea proposed in 1979 by Tajima and Dawson~\cite{tajima_dawson} to use the electric fields generated in a plasma to accelerate electrons has stimulated a broad and rapidly growing research aiming to reduce the typical sizes of the accelerating structures down to the centimetre-scale and develop futuristic table-top machines.
With respect to conventional RF technology, limited to low gradients by electric breakdown, plasma-based devices can sustain fields orders of magnitude larger, up to tens or hundreds of GV/m. So far several schemes have been proposed and realized by employing high-intensity laser pulses~\cite{faure2006controlled,steinke2016multistage} or high-current electron~\cite{2007Natur.445..741B,litos2014high} and proton~\cite{adli2018acceleration} beams to excite very large waves in confined plasma structures. 

The Plasma Wakefield Accelerator (PWFA) concept~\cite{chen1985acceleration} exploits the electric fields of a plasma wave driven by a relativistic electron beam acting as \textit{driver}. Such fields can then be efficiently transferred to a trailing bunch, the \textit{witness}, gaining part of the energy deposited by the driver~\cite{loisch2018observation,roussel2020single}.
Moreover, when driven by ultra-relativistic particle beams, the PWFA is not limited by diffraction nor dephasing~\cite{sprangle1996laser}, making possible large acceleration lengths.
After the first proof-of-principle experiments, further studies were conducted to improve the quality (in terms of energy spread) of the accelerated particle beams~\cite{litos2014high,deng2019generation}.
However, despite several major advances, achieving an energy spread below the percent level has remained an open issue. This problem is of paramount importance for user-oriented applications based, for instance, on Free-Electron Lasers (FELs) which require relative energy spreads well below $1\%$.

\begin{figure*}[t]
\centering
\includegraphics[width=0.95\linewidth]{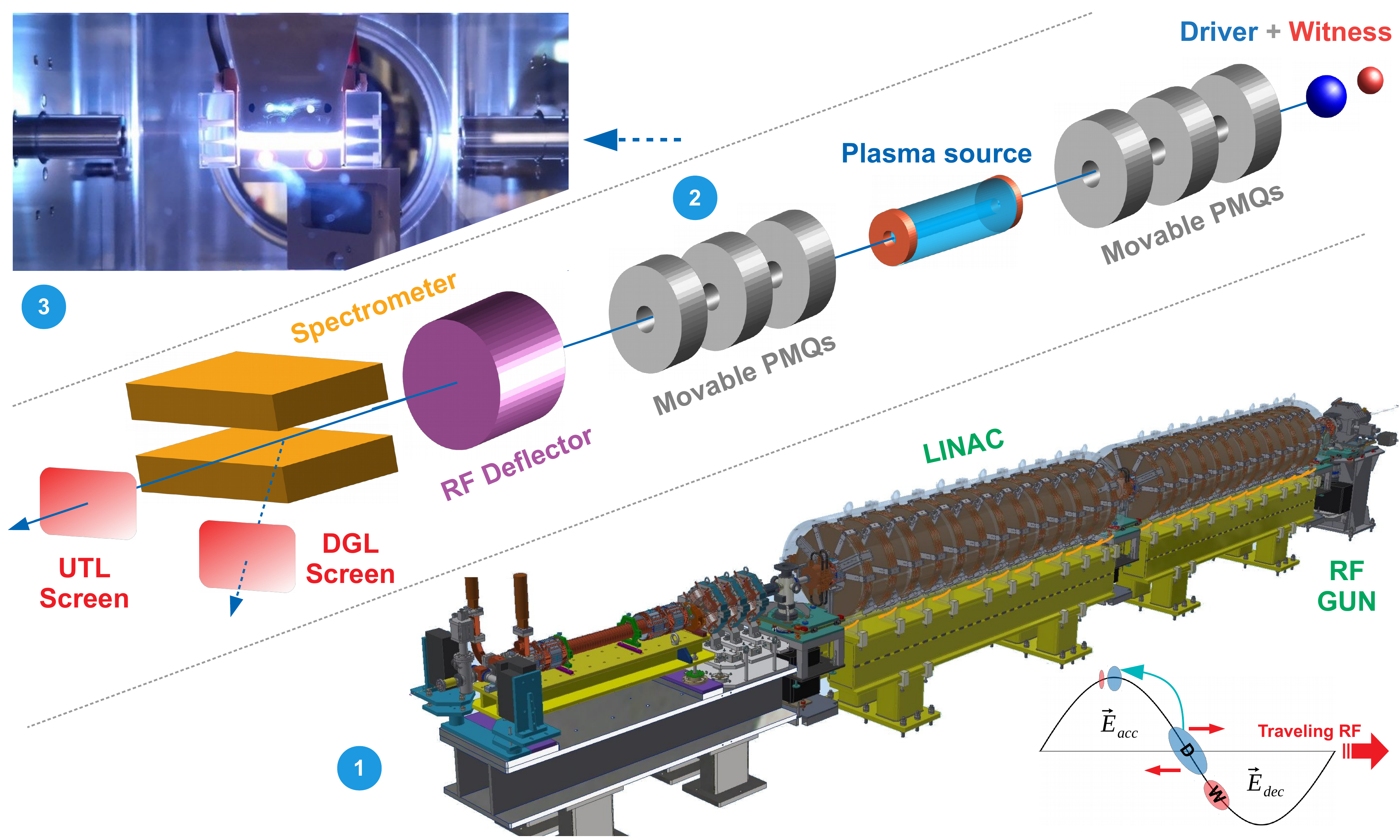}
\caption{Experimental setup. (1) The SPARC photo-injector consists of a radio-frequency (RF) gun followed by a linac with three travelling-wave sections. The driver and witness bunches are produced directly from cathode with two delayed laser pulses. They are then compressed by velocity-bunching in the first section and (2) focused by a triplet of permanent-magnet quadrupoles (PMQs) in the plasma. A second triplet of PMQs is used to extract and transport the bunches in order to be characterized by a magnetic spectrometer and a RF-Deflector. (3) The plasma is produced by ionizing the Hydrogen gas confined in a 3~cm-long capillary by applying a high-voltage discharge to the two electrodes.}
\label{CapillarySetup}
\end{figure*}

Here we report, for the first time, experimental measurements demonstrating the acceleration of a witness bunch with ultra-low energy spread, ultra-short duration (tens of femtoseconds) and tiny transverse size (few microns). We show that, by imprinting a positive energy-chirp on the witness, the PWFA process is exploited to both accelerate and reduce its energy spread. Such a unique feature produced a relative spread as small as $0.1\%$, representing so far the best result achieved for a plasma-based accelerator.

The experiment has been performed at the SPARC\_LAB test-facility~\cite{ferrario2013sparc_lab} by employing two bunches, driver and witness, interacting with the plasma confined in a 3~cm long discharge-capillary~\cite{pompili2018focusing}. Our results show that the witness gains up to 7~MeV in the plasma and, during the acceleration, its longitudinal phase-space (LPS) is rotated and energy spread reduced by $40\%$ with respect to the initial one. 
The energy-chirp imprinted on the witness is thus used to both improve its beam-loading~\cite{tzoufras2008beam} and compensate the slope of the plasma wakefield~\cite{shpakov2019longitudinal,wu2019phase}, so that it can be considered as an \textit{assisted} beam-loading energy spread compensation.
Details about the experimental setup and measurements are given in the following. The study is finally supported by a start-to-end simulation.  

The experimental setup is shown in Fig.~\ref{CapillarySetup}. The electron bunches are produced by the SPARC photo-injector, consisting of a radio-frequency (RF) gun followed by three accelerating sections.
The compression of the two bunches is achieved in the first accelerating section through the velocity-bunching (VB) process~\cite{serafini2001velocity} that slightly decelerates their heads while the tails are accelerated, leading to a compression down to ultra-short durations. 
Solenoid coils around the first section allow to preserve the beam emittance during the process~\cite{ferrario2010experimental}.
The advantage of VB is that acceleration and compression are achieved simultaneously, making the photo-injector very compact. Moreover it allows to precisely adjust the bunch durations, their distance and, with respect to other methods employing masks or scrapers in dispersive sections~\cite{litos2014high,roussel2020single}, there is no loss of charge.
The beam diagnostics consists of a RF-Deflector and a magnetic spectrometer that allow to characterize the temporal profile and energy spectrum of the beam in correspondence of two Ce:YAG scintillator screens located on the straight line (UTL) and on a $14\degree$ dogleg line (DGL)~\cite{cianchi2015six}.

The plasma accelerator module consists of a 3D-printed capillary with 3~cm length and 1~mm diameter (see Fig.~\ref{CapillarySetup}). The capillary is filled at 1~Hz rate with H$_2$ gas through one inlet and has two electrodes at its extremities connected to a high-voltage (HV) discharge generator providing 12~kV pulses with 310~A current.
The plasma density is monitored with a Stark broadening-based diagnostics measuring the H$_{\beta}$ Balmer line~\cite{pompili2018focusing}.
The capillary is installed in a vacuum chamber directly connected with a windowless, three-stage differential pumping system that ensures $10^{-8}$~mbar pressure in the RF linac while flowing H$_2$. This solution allows to transport the beam without encountering any window, thus not degrading its emittance by multiple scattering.
A triplet of movable permanent-magnet quadrupoles (PMQ) is installed upstream the capillary to focus the beam at its entrance~\cite{pompili2018compact}. A second one is put downstream to extract the beam after the acceleration.
The experimental setup is completed by two beam current monitors to measure the beam charge before and after the capillary.

\begin{figure}[h]
\centering
\includegraphics[width=1.0\linewidth]{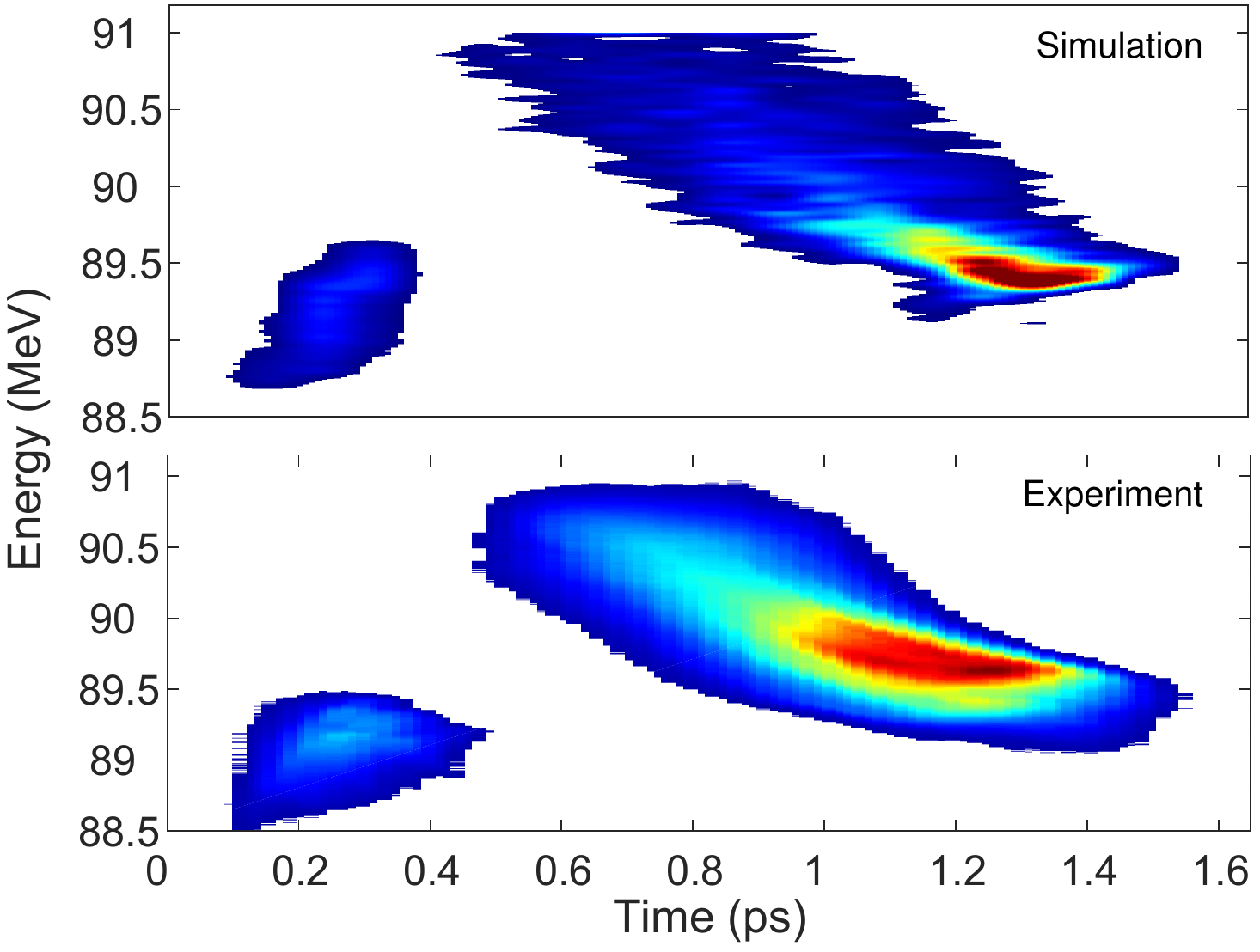}
\caption{Beam configuration. Simulated (top) and experimental (bottom) longitudinal phase-space of the 200~pC driver and 20~pC witness bunches as obtained on the magnetic spectrometer when plasma is turned off.}
\label{LPScomparison}
\end{figure}

The two bunches are generated with the laser-comb technique~\cite{2011NIMPA.637S..43F} by sending two delayed UV pulses on the cathode. The laser energy of each pulse is independently adjusted to produce the required charges.
To avoid the spoiling of the energy spread during the PWFA process, the bunch length is compressed with VB and made much shorter than the wavelength of the accelerating field. 
When operating with plasma densities $n_p\approx 10^{15}\div10^{16}$~cm$^{-3}$ we thus require a witness length of few microns (corresponding to few tens of femtoseconds).
Figure~\ref{LPScomparison} shows the final LPS of the two bunches as measured on the DGL screen with the magnetic spectrometer and RF-Deflector turned on.
The driver has 200~pC charge and 230~fs (rms) duration while the witness has 20~pC charge and 40~fs (rms) duration, corresponding to $\approx 500$~A peak current.
The time distance between the bunches is 1.1~ps.
The figure also shows a numerical simulation reproducing the expected LPS.
The bunches have opposite energy-chirps: negative for the driver (low-energy particles on the head) and positive for the witness (low-energy particles on the tail).
The driver energy is $89.5\pm 0.1$~MeV with $0.25\pm 0.02$~MeV (rms) spread and $2.5\pm 0.1~\mu m$ (rms) normalized emittance.
For the witness the central energy is $E_w= 89.1\pm0.1$~MeV, the spread $\sigma_E= 0.19\pm 0.01$~MeV (corresponding to a relative spread $\sigma_E/E_w\approx 0.2\%$) and normalized emittance $0.9\pm 0.1~\mu m$.

\begin{figure}[!h]
\centering
\subfigure{
\begin{overpic}[width=0.95\linewidth]{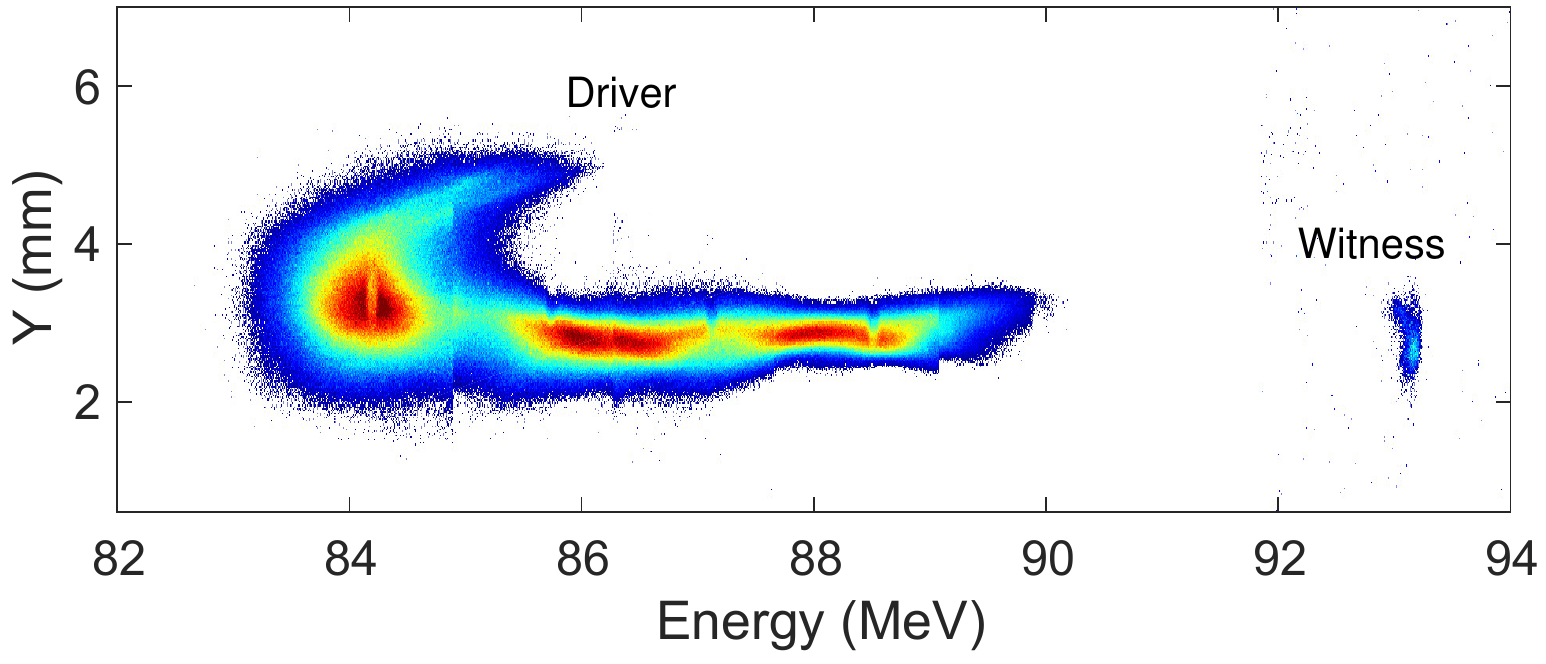}
\put(10,38){\color{black}\textbf{a}}
\end{overpic}
\label{D200W20spectrum}
}
\subfigure{
\begin{overpic}[width=0.95\linewidth]{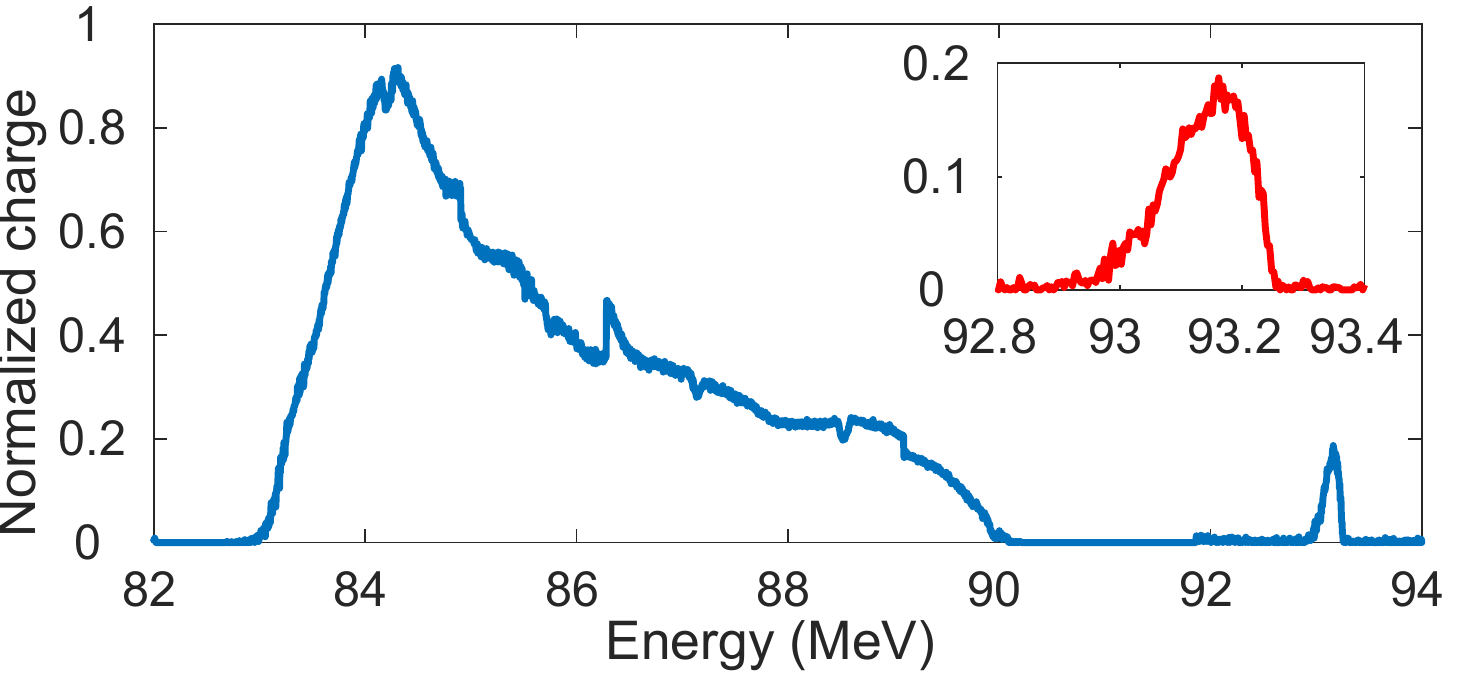}
\put(13,40){\color{black}\textbf{b}}
\end{overpic}
\label{D200W20spectrumProf}
}
\subfigure{
\begin{overpic}[width=0.96\linewidth]{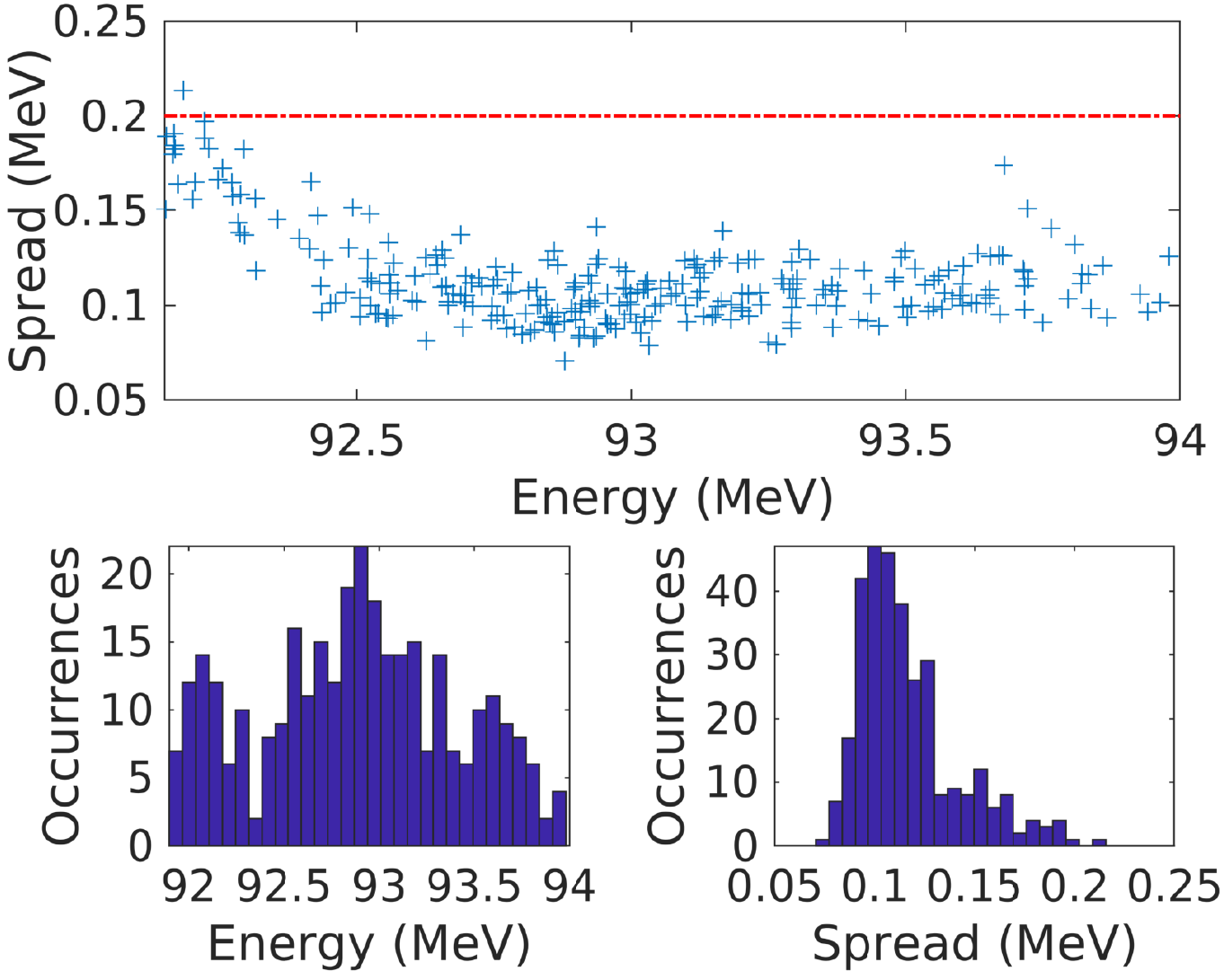}
\put(15,74){\color{black}\textbf{c}}
\end{overpic}
\label{D200W20stat}
}
\caption{Acceleration with the 200~pC driver. (a) Spectrometer image and (b) energy projection. The inset refers to the witness spectrum. (c) Analysis of 320 consecutive shots. The upper plot shows the correlation between the witness central energy and energy spread. The red dashed line is the witness energy spread with plasma turned off. The histograms report the energy and spread distributions.}
\label{D200W20data}
\end{figure}

\begin{figure*}[t]
\centering
\subfigure{
\begin{overpic}[width=0.98\linewidth]{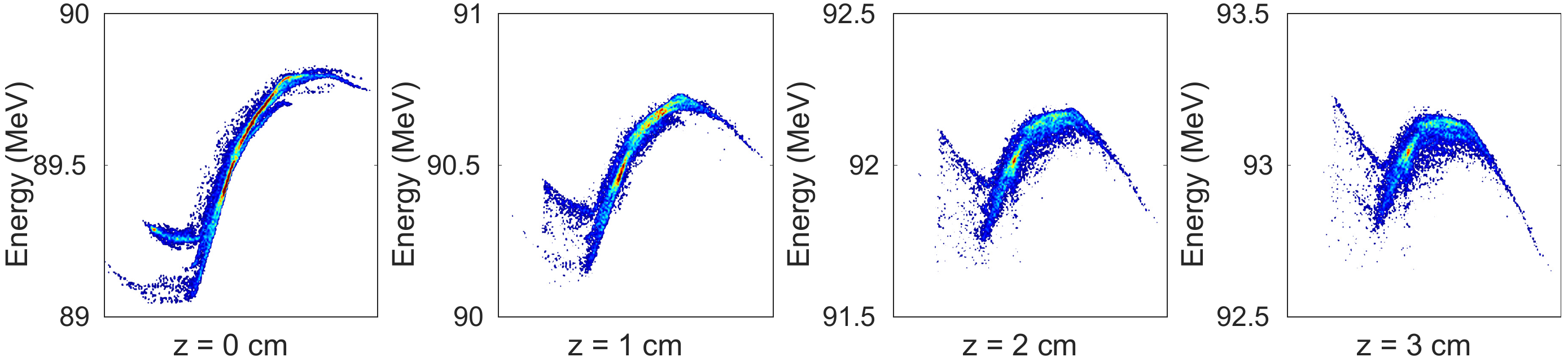}
\put(8,20){\color{black}\textbf{a}}
\end{overpic}
\label{witLPSrot}
}
\subfigure{
\begin{overpic}[height=0.36\linewidth]{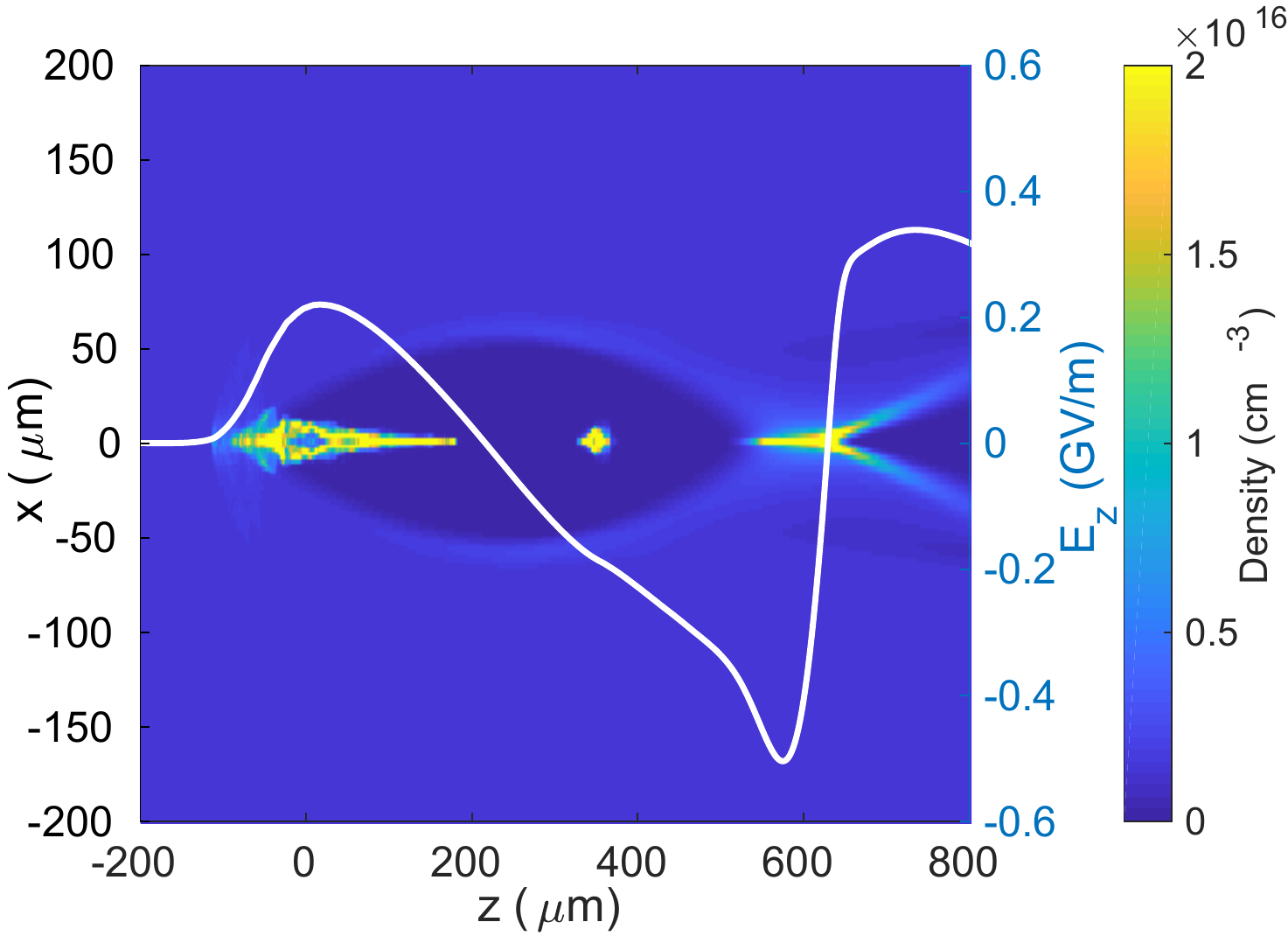}
\put(14,62){\color{white}\textbf{b}}
\end{overpic}
\label{DensityMapArch}
}
\subfigure{
\begin{overpic}[height=0.34\linewidth]{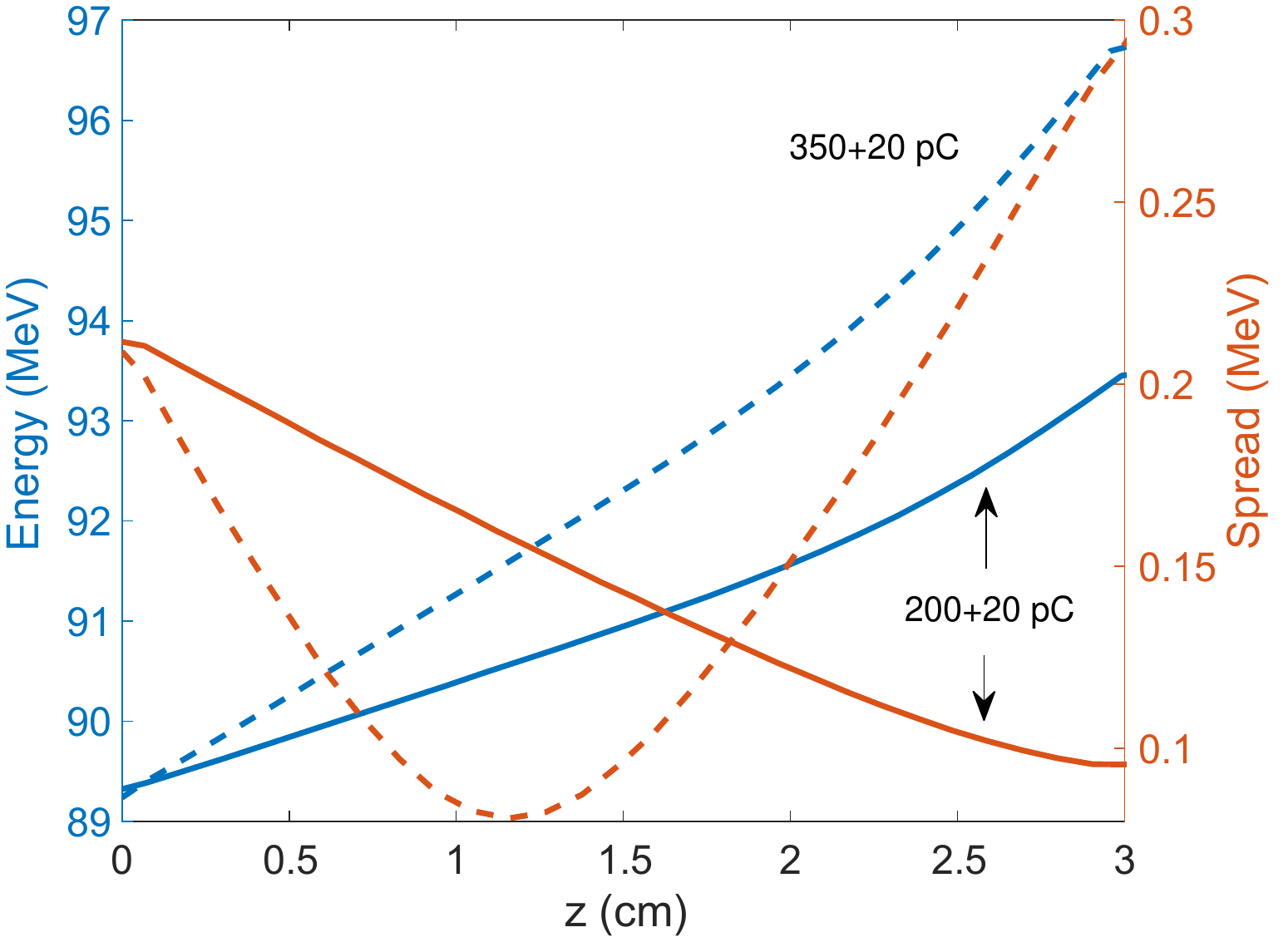}
\put(13,68){\color{black}\textbf{c}}
\end{overpic}
\label{PlaSimAvg}
}
\caption{Plasma wakefield simulations of the beam-driven plasma wakefield acceleration. (a-b) Results for the 200~pC driver configuration. (a) Witness LPS at several positions along the plasma. The frames show the rotation of the LPS leading to the energy spread minimization. (b) Snapshot of the two bunches propagating in the plasma background. The white line show the accelerating field along the axis of symmetry. (c) Evolution of the witness energy (blue) and energy-spread (red) along the 3~cm-long plasma when using the 200~pC (solid line) and 350~pC driver (dashed line) configurations.}
\label{sim_data}
\end{figure*}

To demonstrate the proposed scheme, the beam is focused by the first triplet of PMQs down to $20~\mu m$ at the capillary entrance.
Figure~\ref{D200W20spectrum} shows the reconstructed energy spectrum of the two bunches when the plasma is turned on. The measurement is done on the DGL screen, having an energy acceptance of about 2~MeV. The plasma density is set to $n_p\approx 2\times 10^{15}$~cm$^{-3}$ and can be adjusted by delaying the beam arrival time with respect to the HV discharge pulse.
The driver, when decelerated by the plasma wakefield, spans an energy range of about 7~MeV. To reconstruct its energy spectrum we merged several (partial) spectra acquired for different currents of the magnetic spectrometer. Figure~\ref{D200W20spectrumProf} shows the projection of the energy spectrum, with the inset representing the one relative to the accelerated witness.
To evaluate the overall stability of the PWFA process we acquired 320 consecutive shots.
Figure~\ref{D200W20stat} highlights a resulting mean energy $E_w\pm \Delta E_w=93.1\pm0.5$~MeV and an average accelerating gradient $E_z=133\pm16$~MV/m. The energy stability, defined as $1-\Delta E_w/E_w$, is about $99.5\%$. The most interesting result, however, is the achieved energy spread. It is $\sigma_E=0.12\pm0.03$~MeV (corresponding to a relative spread $\sigma_E/E_w\approx 0.1\%$), thus $40\%$ lower than the one with plasma turned off.
Such a result is attributed to the rotation of the witness LPS during the acceleration so that its correlated energy-chirp is removed and spread minimized.
It can be envisioned by noting that the plasma wakefield is stronger on the lower energy tail of the witness and weaker on its higher energy head, see Fig.~\ref{DensityMapArch}. The head-tail energy deviation is thus reduced accordingly.

\begin{figure}[!b]
\centering
\subfigure{
\begin{overpic}[width=0.95\linewidth]{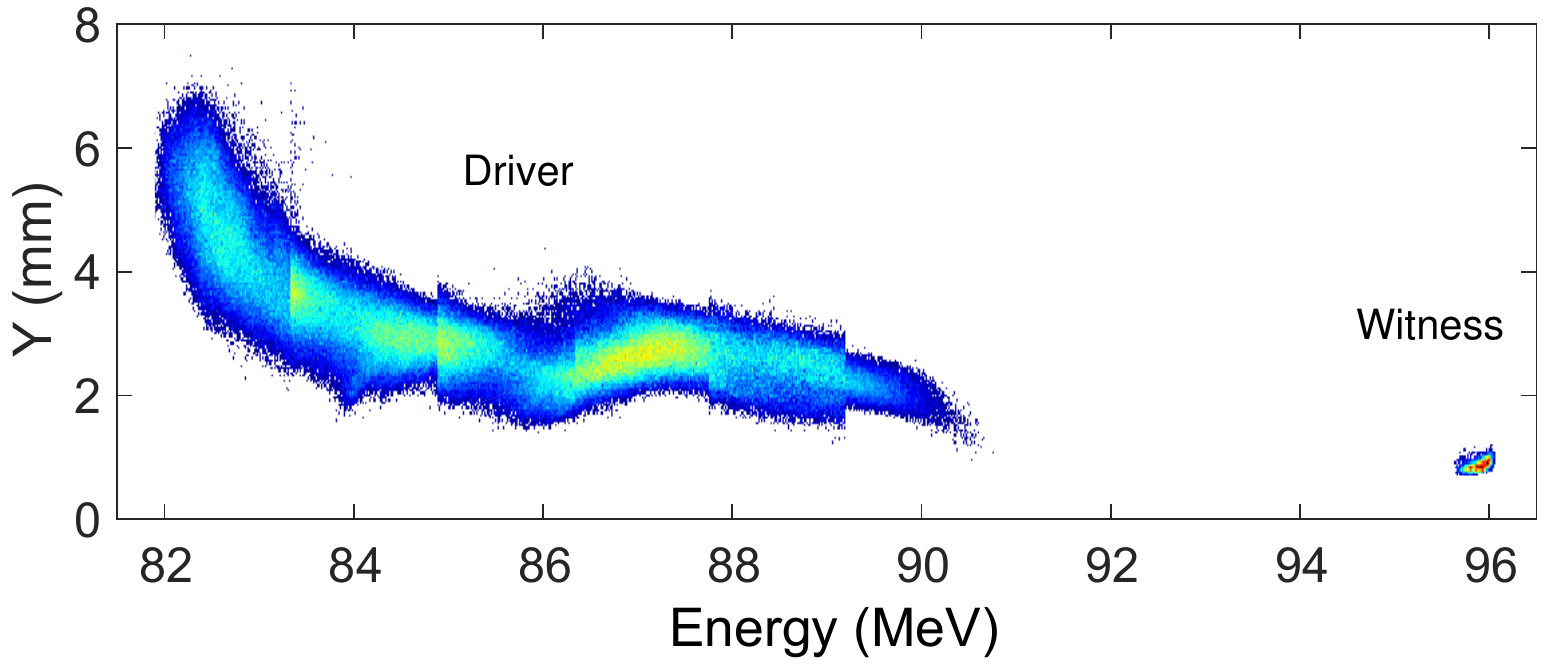}
\put(10,13){\color{black}\textbf{a}}
\end{overpic}
\label{D350W20spectrum}
}
\subfigure{
\begin{overpic}[width=0.95\linewidth]{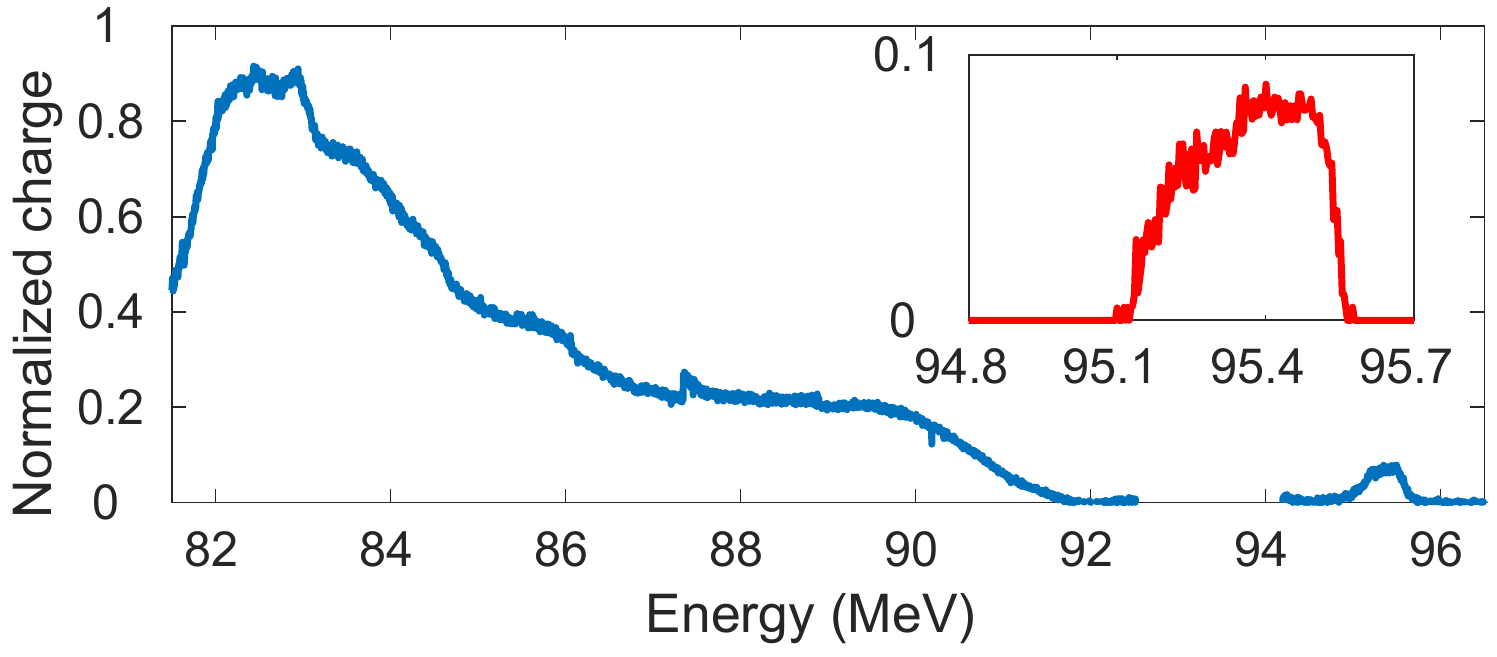}
\put(13,13){\color{black}\textbf{b}}
\end{overpic}
\label{D350W20spectrumProf}
}
\subfigure{
\begin{overpic}[width=0.96\linewidth]{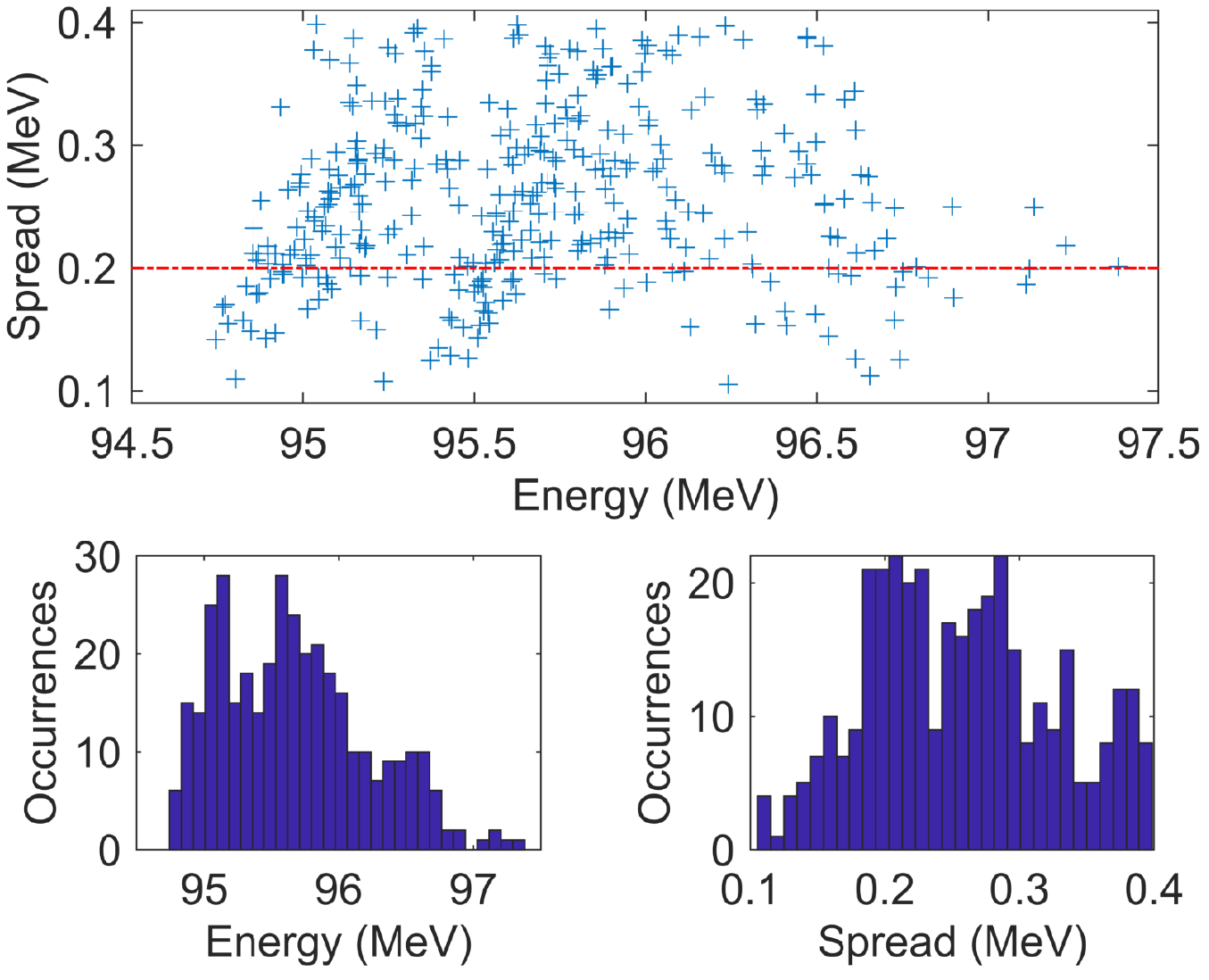}
\put(14,75){\color{black}\textbf{c}}
\end{overpic}
\label{D350W20stat}
}
\caption{Acceleration with the 350~pC driver. (a) Spectrometer image and (b) energy projection. The inset refers to the witness spectrum. (c) Analysis of 360 consecutive shots. The upper plot shows the correlation between the witness central energy and energy spread. The red dashed line is the witness energy spread with plasma turned off. The histograms report the energy and spread distributions.}
\label{D350W20data}
\end{figure}


To extend the previous results and demonstrate the possibility to control and reconstruct the witness LPS during the PWFA process, we increased the driver charge up to 350~pC to generate a larger wakefield and produce an over-rotation of the LPS.
Figure~\ref{D350W20spectrum} shows the reconstructed energy spectrum of the two bunches when the plasma is turned on. The energy-depleted driver now spans a larger energy range ($\approx 8$~MeV) due to the larger decelerating wakefield.
The corresponding energy profile is shown in Fig.~\ref{D350W20spectrumProf}.
The statistical analysis of 360 consecutive shots, reported in Fig.~\ref{D350W20stat}, shows a resulting mean energy of $E_w\pm \Delta E_w=95.9\pm0.6$~MeV, corresponding to an overall energy stability of $99.4\%$ and an average accelerating gradient $E_z=233\pm20$~MV/m. The final energy-spread is now $\sigma_E=0.26\pm0.07$~MeV, i.e. approximately $30\%$ larger than the one with plasma turned off.
As expected the larger acceleration brings to an over-rotation of the witness LPS and, in turn, to a larger spread at the end. It is worth pointing out that the relative spread is however kept very small, $\sigma_E/E_w\approx 0.3\%$, i.e. again an order of magnitude smaller than what achieved in previous plasma wakefield acceleration experiments.

To support the experimental observations, we performed a complete start-to-end simulation where we propagated the two bunches in a 3~cm-long plasma.
The longitudinal plasma profile is numerically described by assuming a plasma density $n_p= 2\times 10^{15}$~cm$^{-3}$ in the central part with decreasing exponential tails extending toward the two openings.
We used the Architect hybrid-kinetic fluid code~\cite{marocchino2016efficient} that simulates the beam particles with a Particle-In-Cell method while the plasma background is treated as a fluid.
The simulation is compared with the experimental observations and, in particular, focuses on the 200~pC driver configuration where the minimum energy spread was achieved.
This scenario is particularly effective since the witness beam-loading is not sufficient to completely counteract the slope of the plasma wakefield and minimize the spread~\cite{tzoufras2008beam}. For this reason the energy-chirp imprinted on the witness is exploited to \textit{assist} the process of energy spread compensation.

The evolution of the witness LPS during its propagation through the plasma is shown in Fig.~\ref{witLPSrot}. Each frame is obtained at different positions along \textit{z}. The plots reveal the progressive rotation of the LPS driven by the plasma wakefield which leads to the acceleration and, at the same time, minimization of the energy spread.
Figure~\ref{DensityMapArch} shows a snapshot at $z=2$~cm of the driver-witness beam propagating in the plasma background. The driver undergoes several transverse modulations due to the radial wakefields while the witness is located in the accelerating region and is pushed to a final energy of 93~MeV, gaining approximately 4~MeV. 
The evolutions of the witness energy and spread along the plasma are shown in Fig.~\ref{PlaSimAvg}, resulting in an excellent agreement with the experiment.
The same applies to the 350~pC driver configuration. The witness energy gain is now larger (about 7~MeV) and, as a consequence, the LPS is over-rotated. The energy spread attains its minimum at $z\approx 1$~cm and then starts to grow, reaching a larger value at the end ($z=3$~cm)

In conclusion, we developed an innovative approach for the minimization of the energy spread from a beam-driven plasma wakefield accelerator. We performed a proof-of-principle experiment where we observed, for the first time, the simultaneous acceleration of the witness and reduction of its energy spread by approximately $40\%$.
The results provide a unique and innovative method to manipulate the witness longitudinal-phase space and optimize its interaction with the plasma. We found that large accelerations can be achieved and energy spread minimized by properly setting a positive energy-chirp on the witness bunch. This method allows to preserve the beam quality during the acceleration and reach energy spreads as low as $0.1\%$, approximately an order of magnitude smaller with respect to what obtained in plasma acceleration experiments so far.
It is worth pointing out that the generation of low energy spread beams is a key requirement when dealing with plasma-based technology. These results represent thus a fundamental step toward the development of next-generation accelerators and demonstrate their effective usability in new compact machines for user-oriented applications.

\appendix
\section*{Acknowledgments}

\begin{acknowledgments}
This work has been partially supported by the EU Commission in the Seventh Framework Program, Grant Agreement 312453-EuCARD-2, the European Union Horizon 2020 research and innovation program, Grant Agreement No. 653782 (EuPRAXIA) and the INFN with the GRANT73/PLADIP grant. The work of one of us (A.Z.) was partially supported by ISF foundation.
The authors thank D. Pellegrini for the realization of the HV discharge pulser and M. Del Franco for providing the layout of the SPARC\_LAB photo-injector.
\end{acknowledgments}

\section*{Author Contributions}
M.F. and R.P. planned and managed the experiment with inputs from all the co-authors.
R.P. carried out the data analysis.
A.B. provided the plasma characterization.
A.C and V.S. realized and managed the beam diagnostics.
A.D.D. provided numerical simulations for the beam-plasma interaction.
R.P. and A.Z. wrote the manuscript.
All authors were involved during the setup and operations of the experiment.
All authors extensively discussed the results and reviewed the manuscript.

\section*{Competing Interests}
The authors declare that they have no competing financial interests.

\section*{Methods}

\subsection*{Generation of two-bunch beam structure}
The driver and witness bunches are directly generated on the photo-cathode by two UV laser pulses delayed by 5.3~ps. The 20~pC witness is emitted before the 200~pC (or 350~pC) driver.
Their charge density is made equal on the cathode to ensure the same space-charge effects and envelope evolution along the linac.
The relative separation is tuned with the velocity-bunching (VB) process and, simultaneously, their durations are compressed. This leads to the final longitudinal phase-space shown in Fig.~\ref{LPScomparison}.
The VB requires the injection in the first travelling-wave section (S1) at the zero-crossing of the RF wave. The longitudinal compression is achieved in the first half of S1, where the tail (head) of the bunch is accelerated (decelerated). The VB process is based on a correlated beam time-velocity chirp, i.e. electrons on the tail of the bunch are made faster than the ones in the head. This leads to a rotation of the LPS if the injected beam is slower than the phase velocity of the RF wave so that when injected at the zero-crossing field phase it slips back to phases where the field is accelerating and is simultaneously chirped and compressed as showed in Fig.~\ref{CapillarySetup}.
To avoid an excessive emittance growth, the S1 embedded solenoids are turned on to provide the necessary additional focusing.
The time-energy-chirps of the resulting bunches are thus complementary because the witness undergoes an over-compression (head and tail are swapped) while the driver is under-compressed. This configuration allowed us to use the driver-excited plasma wakefield to remove part of the witness energy-chirp and reduce its final energy spread accordingly. 

\subsection*{Plasma source and characterization}
For the experiment, a 3~cm-long capillary with 1~mm hole diameter has been used. The capillary is 3D-printed by using photo-polymeric material. It allows to properly shape the capillary and the inlets for the gas injection. The plasma is then produced by ionizing the Hydrogen gas. The Hydrogen is produced by water electrolysis with a NM Plus Hydrogen Generator provided by Linde. The inlet pressure of the gas is set to approximately 200~mbar by a regulator located outside the vacuum chamber. A high speed solenoid valve, located 5~cm far from the capillary, is used to fill the capillary with the gas. The valve stays opened for 3~ms and the discharge-current is applied 1~ms after its closure. The short time of the valve opening is mainly dictated by vacuum requirements of the closest accelerating section, that needs an ultra-high vacuum environment ($~\approx10^{-8}$~mbar). The discharge-current, applied to the two capillary electrodes, is generated by a high-voltage generator providing 12~kV pulses and 310~A current through the capillary. The value of the current is monitored with a Pearson current monitor placed around one of the two wires connected with the capillary electrodes.
The plasma density is retrieved with a Stark broadening-based diagnostics by measuring the $H_{\alpha,\beta}$ Balmer lines.

\subsection*{Photo-injector and plasma wakefield acceleration simulations}
The start-to-end simulation is performed by means of two simulation codes. The beam dynamics along the SPARC photo-injector, from its generation and up to the injection into the plasma, is analyzed with the General Particle Tracer (GPT) code~\cite{gpt_web}, widely used in the accelerator community. The code simulates all the elements of the photo-injector and includes space-charge effects. The driver and witness bunches contain $7.5\times 10^5$ and $2.5\times 10^5$ macro-particles, respectively. The simulated data is cross-checked with the experimental measurements by means of several Ce:YAG scintillator screens installed along the machine.
The interaction of the bunches with the plasma background is simulated with Architect~\cite{marocchino2016efficient}. The code uses an hybrid approach; the beam particles are treated kinetically as in a Particle-In-Cell (PIC) code while the plasma background is modelled as a fluid. The code uses the two bunches simulated by GPT as input. The simulation described here has been obtained with $2~\mu m$ mesh resolution in the longitudinal and transverse dimensions. The plasma density profile is computed in order to reproduce the one experimentally measured.

\bibliography{biblio}
\bibliographystyle{apsrev4-1}

\end{document}